\documentclass[12pt,a4paper]{article}
\pdfoutput=1
\usepackage[utf8]{inputenc}
\usepackage[T1]{fontenc}
\usepackage[swedish,english]{babel}
\usepackage{lmodern}
\usepackage[table]{xcolor}
\usepackage{graphicx}
\usepackage[intlimits]{amsmath}
\usepackage{icomma} 
\usepackage{bbm} 
\usepackage{amssymb}
\usepackage{cases} 

\usepackage{a4wide}

\usepackage{mathtools} 
\usepackage{bm} 

\usepackage[normalem]{ulem} 
\usepackage{textcomp} 
\usepackage{footnote} 
\usepackage{pdfpages} 
\usepackage[font=small, labelfont=bf]{caption}  
\usepackage{pbox} 

\usepackage{subcaption}	

\usepackage{bbold}	

\definecolor{plotblue}{RGB}{94,129,181}
\definecolor{darkplotblue}{RGB}{47,64,90}
\usepackage[colorlinks=true,linkcolor=darkplotblue,citecolor=black,urlcolor=black]{hyperref}

\definecolor{gray}{gray}{.6}
\definecolor{lightgray}{gray}{.8}
\definecolor{almostwhite}{gray}{.99}

\newcommand{\tskip}[1]{\vphantom{\rule{0 mm}{#1 em}}}

\newcommand{\N}{\ensuremath{\mathbbm{N}}}

\newcommand{\R}{\ensuremath{\mathbbm{R}}}

\DeclareMathOperator{\tr}{tr} 

\newcommand{\defeq}{\vcentcolon=} 
\newcommand{\eqdef}{=\vcentcolon} 
\newcommand{\st}[1][]{%
  \ifthenelse{\isempty{#1}}{\ensuremath{\;|\;}}%
  {\ensuremath{\;#1|\;}}}	

\newcommand{\rd}{\ensuremath{\mathrm{d}}}
\newcommand{\id}{\ensuremath{\!\rd}}	
\newcommand{\e}{\ensuremath{\mathrm{e}}}	
\newcommand{\I}{\ensuremath{\mathrm{i}}}	
\newcommand{\tfh}{\ensuremath{\tfrac{1}{2}}}	
\newcommand{\eye}{\ensuremath{\mathbbm{1}}}	
\DeclarePairedDelimiter\bra{\langle}{\rvert}
\DeclarePairedDelimiter\ket{\lvert}{\rangle}
\DeclarePairedDelimiterX\braket[2]{\langle}{\rangle}{#1 \delimsize\vert #2}
\DeclarePairedDelimiter\expval{\langle}{\rangle}

\renewcommand{\vec}[1]{\ensuremath{\mathbf{#1}}}
\newcommand{\liegroup}[1]{\ensuremath{#1}}	
\newcommand{\U}{\liegroup{U}}			
\newcommand{\SU}{\liegroup{SU}}			
\newcommand{\SO}{\liegroup{SO}}			

\newcommand{\op}{\ensuremath{\mathcal{O}}}

\newcommand{\MPS}{\text{MPS}}
\newcommand{\bMPS}{\ensuremath{\bra{\text{MPS}}}}
\newcommand{\kMPS}{\ensuremath{\ket{\text{MPS}}}}
\newcommand{\ua}{\ensuremath{\uparrow}}
\newcommand{\da}{\ensuremath{\downarrow}}
\newcommand{\perm}{\ensuremath{\mathbb{P}}}  
\newcommand{\parity}{\ensuremath{\Pi}}  	
\newcommand{\vac}{\ensuremath{\Omega}}	
\newcommand{\kvac}{\ensuremath{\ket{\vac}}}	
\newcommand{\vn}{\ensuremath{\vec{n}}}
\newcommand{\vs}{\ensuremath{\vec{s}}}

\newcommand{\transf}{\ensuremath{t}}		

\newcommand{\Xb}{\ensuremath{\overline{X}}}
\newcommand{\Yb}{\ensuremath{\overline{Y}}}
\newcommand{\Zb}{\ensuremath{\overline{Z}}}
\newcommand{\dagPhi}{\ensuremath{\Phi^{\dagger}}}

\newcommand{\Fb}{\ensuremath{\bar{F}}}
\newcommand{\rscal}{\ensuremath{\varphi}}		
\newcommand{\cscal}{\ensuremath{\phi}}		
\newcommand{\cscalb}{\ensuremath{\bar{\phi}}}		

\newcommand{\cl}{\ensuremath{\text{cl}}}	
\newcommand{\tree}{\ensuremath{\text{tree}}}	

\newcommand{\thetab}{\ensuremath{\bar{\theta}}}
\newcommand{\lambdab}{\ensuremath{\bar{\lambda}}}
\newcommand{\alphad}{\ensuremath{\dot{\alpha}}}

\newcommand{\Wb}{\ensuremath{\overline{W}}}

\newcommand{\comm}[3][]{{}[{}#2{}\,{\overset{#1}{,}}\,{}#3{}]{}}

\newcommand{\cstar}{\ensuremath{\ast}}
\newcommand{\starcomm}[2]{\comm{#1}{#2}_\cstar}

\newcommand{\q}{\ensuremath{\mathbf{q}}}

\newcommand{\hc}{\ensuremath{\text{h.c.}}}
\newcommand{\bt}{\ensuremath{\text{b.t.}}}

\newcommand{\fatZero}{\ensuremath{\mathbb{0}}}

\newcommand{\lag}{\ensuremath{\mathcal{L}}}

\DeclarePairedDelimiter\floor{\lfloor}{\rfloor}

\begin{document}

\begin{titlepage}

\begin{flushright}\footnotesize
  \texttt{NORDITA-2018-034} \\
  \texttt{UUITP-17/18}
\vspace{0.3cm}
\end{flushright}

\centerline{\large \bf One-point functions in $\beta$-deformed $\mathcal{N}=4$ SYM with defect}
\vskip 0.7 cm

\centerline{\bf Erik Wid\'en}

\vskip 0.6cm

\begin{center}
  \small \sl
  NORDITA\\
    KTH Royal Institute of Technology and Stockholm University\\
    Roslagstullsbacken 23, SE-106 91 Stockholm, Sweden\\[1mm]
  Department of Physics and Astronomy, Uppsala University\\
  SE-751 08 Uppsala, Sweden\\[1mm]
  \texttt{\upshape erik.widen@nordita.org}
\end{center}
\vskip 0.7cm

\begin{abstract}
  We generalize earlier results on one-point functions in $\mathcal{N}=4$ SYM with a co-dimension one defect, dual to the D3-D5-brane setup in type IIB string theory on $AdS_5\times S^5$, to a similar setup in the $\beta$-deformed version of the theory. The treelevel vacuum expectation values of single-trace operators in the two-scalar-subsector are expressed as overlaps between a matrix product state (MPS) and Bethe states in the corresponding twisted spin-chain picture. We comment on the properties of this MPS and present the simplest analytical overlaps and their behavior in a certain limit (of large $k$). Importantly, we note that the deformation alters earlier interpretations of the MPS as an integrable boundary state, seemingly obstructing simplifications of the overlaps analogous to the compact determinant formula found in the non-deformed theory. The results are supplemented with some supporting numerical results for operators of length eight with four excitations. 
\end{abstract}

\end{titlepage}

\section{Introduction}
One-point functions in conformal field theories with a defect (dCFTs) and a string theory dual have been studied from an integrability perspective in a series of papers the last years. These are works continuing the now one and a half decade's usage of integrability techniques to explore and extend the powers of holography and the AdS/CFT correspondence. Since the introduction of the spin-chain picture into this context in \cite{ci:Minahan:2002ve}, there has been an extensive amount of work branching out with ever growing patches of understanding of field theories, string theory and the dualities that link them. Reference \cite{ci:Beisert:2010jr} provides a pedagogical overview of the first decade of these endeavours but a lot of progress has been made since.

As stated, the studies of one-point functions form one of these more recent lines of inquires. More than one setting have been in the scope\footnote{See for instance \cite{ci:deLeeuw:2016ofj}.} but the longest string of papers has concerned the defect $\mathcal{N}=4$ SYM dual to the D3-D5 brane system with $k$ units of flux, briefly reviewed below. The first paper to appear, \cite{ci:deLeeuw:2015hxa}, noticed that treelevel $\SU(2)$-subsector one-point functions --- allowed here by the breaking of the conformal symmetry by the defect --- could be written as overlaps between Bethe states and a matrix product state (MPS) in the spin-chain picture. The overlaps could further be simplified into a compact determinant formula. These results were generalized in \cite{ci:BuhlMortensen:2015gfd} and extended to the $\SU(3)$-subsector in \cite{ci:deLeeuw:2016umh}. Recently, the determinant formula was found for the entire scalar $\SO(6)$-subsector in \cite{ci:deLeeuw:2018mkd}. In parallel, efforts have been made to move beyond treelevel beginning with \cite{ci:BuhlMortensen:2016pxs}, later fleshed out in \cite{ci:BuhlMortensen:2016jqo}, which lay the framework for loop computations in the defect $\mathcal{N}=4$ SYM. An asymptotic higher-loop generalization of the determinant formula for the $\SU(2)$-subsector was then proposed in \cite{ci:Buhl-Mortensen:2017ind}. Many of these results are gathered and explained in the lecture notes \cite{ci:deLeeuw:2017cop} from the Les Houches Summer School 2016.

Two-point functions in this setting have also attracted some attention, e.g. \cite{ci:Liendo:2016ymz,ci:deLeeuw:2017dkd,ci:Widen:2017uwh}.\bigskip

\noindent Another and major topic in the integrability approach to AdS/CFT is the study of deformations. The successful techniques in the original $AdS_5\times S^5$ string theory/$\mathcal{N}=4$ SYM version of the correspondence have and continue to be extended to more general pairs of theories by introducing continuous deformations carefully designed to preserve integrability. An early and well studied example of this is the so called $\beta$-deformation, first introduced on the field theory side in \cite{ci:Leigh:1995ep} and later brought to the holography context in \cite{ci:Lunin:2005jy}. It allows for calculations in a slightly modified spin-chain picture where the deformation parameter $\beta$ enters as a twist in the boundary conditions, see \cite{ci:Roiban:2003dw,ci:Berenstein:2004ys,ci:Frolov:2005ty,ci:Beisert:2005if,ci:Gromov:2010dy} among others.\bigskip

\noindent This paper begins to join these two branches by studying a vacuum solution corresponding to a defect in the $\beta$-deformed $\mathcal{N}=4$ SYM, focusing on its two-scalar-subsector treelevel one-point functions by use of the spin-chain formulation. 

The paper is organized as follows. 
\underline{Section \ref{sec:reviewOfNonDeformedCase}} contains a review of the non-deformed theory and the known relevant results.
\underline{Section \ref{sec:reviewDeformed}} introduces the $\beta$-deformation and the corresponding spin-chain picture.
\underline{Section \ref{sec:onepointFuncsInDeformed}} studies the one-point functions in the $\beta$-deformed theory with subsections for the derivation of a classical scalar solution, the spin-chain consequences, some simple analytic examples and, lastly, numerics for a few spin-chain overlaps. The paper is concluded in \underline{section \ref{sec:conlusions}}.

\section{Brief review of the non-deformed case}\label{sec:reviewOfNonDeformedCase}
Non-constant one-point functions are forbidden by translational invariance but they appear in the context of dCFTs due to the defect breaking this symmetry. Building on former works, we begin by briefly reviewing this in the setting of the defect $\mathcal{N}=4$ SYM dual to the D3-D5 system.

On the string theory side, the setup is the so called fuzzy funnel solution to a stack of $N$ D3-branes intersected in three dimensions by a probe D5-brane \cite{ci:Karch:2000gx,ci:NAHM1980413,ci:Diaconescu:1996rk,ci:Constable:1999ac}. A background gauge field has $k$ units of flux through an $S^2$-part of the geometry of the latter, which amounts to $k$ of the $N$ D3-branes dissolving into the D5-brane. This is mirrored in the field theory dual by a symmetry breaking in the form of a co-dimension one defect situated at the field theory coordinate $z=0$. It separates the region $z>0$ with gauge group $\SU(N)$ from the region $z<0$ where the gauge group is broken down to $\SU(N-k)$ by non-zero scalar vacuum expectation values. More specifically, three out of the six real $\mathcal{N}=4$ SYM scalars acquire the classical values
\begin{equation}
  \rscal^{\cl}_{i} = \frac{1}{z} \; t_{i} \oplus \fatZero_{(N-k)}, \quad i= 1, 2, 3,  
  \quad \text{while} \quad
  \rscal^{\cl}_{j} = 0, \quad j= 4, 5, 6
  .
  \label{eq:undefClassSol}
\end{equation}
The matrices $\{t_1, t_2, t_3\}$ form a $k\times k$ unitary representation of $\SU(2)$ and $\fatZero_{(N-k)}$ is the $(N-k)\times (N-k)$ zero matrix (note that we will redefine $t_i$ below).

The field content of the theory enlarges that of ordinary $\mathcal{N}=4$ SYM to also include new fields localized to the defect. They interact both within themselves and with the bulk ($z>0$) fields but will not be of importance here, nor will in fact any fields but the scalars from $\mathcal{N}=4$ SYM. 

The latter group of fields form a subsector of single trace operators closed under the action of the one-loop dilatation operator. Switching to the complex scalar basis 
\begin{align*}
  Z &= \cscal^1 = \rscal_1 + \I \, \rscal_4 \,,  	&	X &= \cscal^2 = \rscal_2 + \I \, \rscal_5 \,, 	&	Y &= \cscal^3 = \rscal_3 + \I \, \rscal_6 \,,
  \\
  \Zb &= \cscalb_1 = \rscal_1 - \I \, \rscal_4 \,,  	&	\Xb &= \cscalb_2 = \rscal_2 - \I \, \rscal_5 \,, 	&	\Yb &= \cscalb_3 = \rscal_3 - \I \, \rscal_6 \,,
\end{align*}
the diagonal operator basis can be written as 
\begin{equation*}
  \op_{\Psi} = \Psi^{i_1 \ldots i_L} \tr \left( X_{i_1} \cdots X_{i_L} \right) , \qquad X_{i} \in \{Z, X, Y, \Zb, \Xb, \Yb \}
  .
\end{equation*}

\paragraph{The spin-chain picture and the $\SU(2)$-subsector.}
The one-loop dilatation operator acts on $\op_{\Psi}$ as does an $\SU(4)$ spin-chain Hamiltonian on a corresponding Hilbert space in which the states are identified with the traces of field products with each field $X_{i_\ell}$ determining the spin at site $\ell$. This and similar spin-chains are well-known, integrable and solvable through various versions of the Bethe ansatz; the coefficients $\Psi^{i_1 \ldots i_L}$ are hence simply the Bethe wavefunctions. 

As mentioned in the introduction, there is now a determinant formula for all one-point functions at treelevel in this subsector \cite{ci:deLeeuw:2018mkd}. We will however be concerned exclusively with the smaller $\SU(2)$-subsector from now on. It consists of operators built only from two of the six scalars --- we choose $Z$ and $X$ --- which map to spin up and spin down in an $XXX_{1/2}$ Heisenberg spin-chain; 
we have the two pictures
\newcommand{\locTskip}{\tskip{1.5}}
\newlength{\arrLineSep}
\setlength{\arrLineSep}{3mm}
\begin{align*}
  \begin{array}[t]{c}
    \text{field theory}\\
    \arrayrulecolor{gray}\hline\locTskip
      \op_{\Psi} = \Psi^{i_1, \ldots, i_{L}} \tr \left( X_{i_1} \cdots X_{i_L} \right)
      \\[\arrLineSep]
      X_1 = Z,  	\qquad X_2 = X,
      \\[\arrLineSep]
      \tr\left( Z Z \cdots Z \right) ,
  \end{array}
  \qquad	
  \begin{array}[t]{c}
    \vphantom{\text{field theory}}\\
    \vphantom{\locTskip}\\[\arrLineSep]
    \longleftrightarrow
  \end{array}
  \qquad
  \begin{array}[t]{c}
    \text{spin-chain}\\
    \arrayrulecolor{gray}\hline\locTskip
	  \Psi^{i_1 \ldots i_L}  \ket{s_{i_1}, \cdots, s_{i_L}} \eqdef \sum_\vs \Psi_{\vs} \ket{\vs}
          \\[\arrLineSep]
          \ket{s_1} = \ket{\ua}, 	\qquad  	\ket{s_2} = \ket{\da},
          \\[\arrLineSep]
          \kvac  = \ket{\ua \ua \cdots \ua} 
	  .
  \end{array}
\end{align*}
Here we also took the chance to introduce the compact notation $\vs$ representing an ordered set of spin variables and to define the spin-chain vacuum $\kvac$. We will refer to spin-downs as excitations and also use the notation $\vn = \{n_1, n_2, \cdots, n_M\}$ for their site numbers such that $\ket{\ua \da} = \ket{s_1, s_2} = \ket{n_1 = 2}$. The total excitation number $M$ is conserved due to the full $\SU(2)$-symmetry.

The matrix of anomalous dimensions $\Gamma$ in this subsector is proportional to the $XXX_{1/2}$ Hamiltonian in the spin-chain picture,
\begin{gather}
  \Gamma = \frac{\lambda}{16 \pi^2} H	\, ,
  \\
  H = 2 \sum_{\ell=1}^L (\eye - \perm_{\ell, \ell +1})
  =  \sum_{\ell=1}^{L} \left(1 - 4 \, S^3_{\ell} S^{3}_{\ell+1} - 2\,  S^{+}_{\ell} S^-_{\ell+1} - 2 \, S^{-}_{\ell} S^{+}_{\ell+1} \right)
   ,
\end{gather}
where $\lambda = g^2 N$ is the 't Hooft coupling, $\perm$ is the permutation of two spin-chain sites and $S^{\pm, 3}_{\ell}$ are the usual spin operators acting at site $\ell$.
It is diagonalized by the (coordinate) Bethe ansatz which is parametrized by $M$ momentum parameters $p_i$ or, equivalently, by the corresponding rapidities $u_i = \tfrac{1}{2} \cot \tfrac{p_i}{2}$. The ansatz reads
\begin{equation}
  \ket{\Psi} =  \sum_{\sigma\in S^M} \sum_{1 \leq n_1 < \ldots < n_M \leq L} 
  S_{\sigma} \; \e^{\I \sum_i p_{\sigma_i} n_i} \; \ket{\vn}
\end{equation}
which includes the product of $S$-matrices $S_{\sigma}$ that corresponds to the permutation $\sigma$, e.g. 
\begin{equation}
  S_{1432} = S_{34} S_{24} S_{23} , \qquad S_{ij} = - \frac{1 + \e^{\I(p_i + p_j)} - 2 \, \e^{\I p_j}}{1 + \e^{\I(p_i + p_j)} - 2\,  \e^{\I p_i}} = \frac{u_i - u_j - \I}{ u_i - u_j + \I}
  .
\end{equation}
These states are Hamiltonian eigenstates, and to all higher charges in the commuting integrability tower of such, provided the rapidities satisfy the Bethe equations
\begin{equation}
  \left( \frac{u_k + \I/2}{u_k - \I/2} \right)^{L}
  \prod_{j\neq k} \frac{u_k-u_j - \I}{u_k - u_j + \I}
  = 1
  ,
\end{equation}
whence they are also referred to as Bethe roots. A Bethe state with rapidities satisfying the Bethe equations is said to be on-shell and in order for it to map to a single-trace operator in the field theory, it has to be translationally invariant since the trace is cyclic.

\paragraph{The determinant formula for the $\SU(2)$-subsector.}
By use of the spin-chain picture, one-point functions can be expressed as an overlap with an MPS designed to produce the correct traces of $t$-matrices. Staying at treelevel in the $\SU(2)$-sector, the MPS reads
\begin{equation}
  \label{eq:defMPS}
  \bMPS = \tr\left( t_1 \bra{\ua} + t_2 \bra{\da} \right)^{\otimes L}
  ,
\end{equation}
such that $\expval{\op_{\Psi}}_{\tree} \propto \braket{\MPS}{\Psi}$. The full formula, valid for any positive integer $k$, is
\begin{align}
  \expval{\op_{\Psi}}_{\tree} = \frac{2^{L-1}}{z^L}
  C_2
\sum_{j=\frac{1-k}{2}}^{\frac{k-1}{2}}  j^L \prod_{i=1}^{\frac{M}{2}} 
\frac{u_i^2\left(u_i^2 + \frac{k^2}{4}\right)}{\left[u_i^2+(j-\tfrac{1}{2})^2\right]
\left[u_i^2+(j+\tfrac{1}{2})^2\right]}
,
\end{align}
where the spin-chain overlap is packed into\footnote{The subscript ${}_{2}$ refers to $C_2$ being the answer for $k=2$ which extends to general $k$ by the given formula.} 
\begin{equation}
  \label{eq:undefOverlap}
  C_2 = \left( \frac{8\pi^2}{\lambda} \right)^{L/2} \! \frac{1}{\sqrt{L}} \,\frac{\braket{\MPS_{k=2}}{\Psi}}{\sqrt{ \braket{\Psi}{\Psi}}}
  .
\end{equation}
In addition to $\ket{\Psi}$ needing to be on-shell, there are further conditions for this formula to describe a non-zero overlap:
\begin{itemize}\itemsep0pt
  \item both $L$ and $M$ need to be even,
  \item all rapidities need to be finite\footnote{%
    Bethe states with finite rapidities are highest weight states in the irreducible decomposition of the $\SU(2)$-representation $\mathbf{2}^{\otimes L}$. Adding rapidities at infinity trivially satisfies the Bethe equations and corresponds to lower weight states called Bethe descendants. These require a slight modification of the formula above, see appendix in \cite{ci:deLeeuw:2017dkd}.
  }
 and
  \item the rapidities need to come in pairs of the form $\{u_1, -u_1, u_2, - u_2, \dots\}$.
\end{itemize}
The first condition is a consequence of the traces of $t$-matrices --- we will return to this below --- while the third is due to annihilation of $\kMPS$ by the third charge $Q_3$. In a quite recent paper on boundary states and integrability, a boundary state such as the MPS was defined as integrable if it is annihilated by all odd charges, i.e.
\begin{equation}
\label{eq:defIntegrableMPS}
  Q_{2n-1} \kMPS = 0, \qquad n\in \N
  ,
\end{equation}
which is implied by the slightly stronger requirement
\begin{equation}
  \parity \; \transf(u) \parity \kMPS = \transf(u) \kMPS
  .
\end{equation}
Here $\parity$ denotes the parity operator which effectively reverses the site ordering in the chain while $\transf(u)$ is the transfer matrix in the algebraic Bethe ansatz formalism\footnote{%
  No review of this will be provided; if needed see for instance \cite{ci:Faddeev:1996iy}
}
 at spectral parameter value $u$.
This definition was devised for theories where
\begin{align}
  \parity Q_{n} \parity = (-1)^{n} Q_n
  ,
\end{align}
which indeed is the case here but will no longer be true in the deformed case below. The integrable MPS with its rapidity pairing grants the drastic simplification of the overlap \eqref{eq:undefOverlap} into the determinant formula
\begin{equation}
  C_2  =2\left[
 \left(\frac{2\pi ^2}{\lambda }\right)^L\frac{1}{L}
 \prod_{j}^{}\frac{u_j^2+\frac{1}{4}}{u_j^2}\,\,\frac{\det G^+}{\det G^-}\right]^{\frac{1}{2}}
 ,
\end{equation} 
where $G^{\pm}$ are $(M/2)\times (M/2)$-dimensional matrices with rapidity dependent matrix elements
\begin{align*}
 G^\pm_{jk}&=\left(\frac{L}{u_j^2+\frac{1}{4}}-\sum_{n}^{}K^+_{jn}\right)\delta _{jk}
 +K^\pm_{jk},
 & \text{with} &&
 K^\pm_{jk}&=\frac{2}{1+\left(u_j-u_k\right)^2}\pm
 \frac{2}{1+\left(u_j+u_k\right)^2}\, .
\end{align*}

This concludes the acount of the relevant results from the non-deformed theory and we move on to review the $\beta$-deformation.

\section[The beta-deformation]{The $\beta$-deformation}\label{sec:reviewDeformed}
The $\beta$-deformation is commonly introduced by starting from the $\mathcal{N}=1$ superspace action of $\mathcal{N}=4$ SYM and replacing all multiplications of fields with a non-commutative $\cstar$-product \cite{ci:Lunin:2005jy,ci:Frolov:2005dj}. 

The $\cstar$-product is expressed in terms of the $SU(4)_R$-charges of the factors according to
\begin{equation}
  \label{eq:starProduct}
  A \cstar B = \e^{\frac{\I}{2} \q_A \wedge \q_b} A B 
  ,
\end{equation}
where the anti-symmetric product of $SU(4)_R$-Cartan charge vectors $\q_A$ and $\q_B$, belonging to operator $A$ and $B$ respectively, is defined through the anti-symmetric matrix of deformations
\begin{align}
  \q_A \wedge \q_B &= \q_A^T \, \mathbf{C} \, \q_B 
  , &
  \mathbf{C} &= \begin{pmatrix}
    0		& -\gamma_3	& \gamma_2	\\
    \gamma_3	& 0		& -\gamma_1	\\
    -\gamma_2	& \gamma_1	& 0
  \end{pmatrix}
    .
\end{align}
The three deformation parameters $\gamma_i$ allow for the more general $\gamma$-deformation, among which the $\beta$-deformation is a special case obtained by setting
\begin{equation}
  \gamma_i = -\beta, 	\qquad	\beta\in \R
  .
\end{equation}
In addition to still being integrable, this special case also preserves one of the supersymmetry generators.

Specifically for the complex scalars, we have the definite charge vectors $[\q_{\cscal^i}]^{\mathrm{c}} = \delta^{\mathrm{c}}_i $ (with a minus sign for the conjugates), with the $SU(4)_R$-Cartan index $c = 1, 2, 3$, and we define
\begin{equation}
  \cscal^1 \cstar \cscal^2 = 
  q \; \cscal^1 \cscal^2, \qquad \qquad q \defeq   \e^{\frac{\I}{2} \beta} 
  .
\end{equation}

Multiple $\cstar$-products are associative and evaluate to
\begin{gather}
  A_1 \cstar A_2 \cstar \dots \cstar A_L = 
  \exp \left[ \tfrac{\I}{2} \sum_{i<j} \q_{A_i}\wedge \q_{A_j} \right] A_1 A_2 \dots A_L
  .
\end{gather}

\paragraph{The $\beta$-deformed action.}
We now follow through with the deformation and insert the $\cstar$-product into the superspace action. 
However, the anti-symmetry of $\wedge$ reduces all field $\cstar$-multiplications to ordinary products except those inside commutators. We denote this as $\starcomm{\cdot}{\cdot}$, which should be understood as regular commutators but with all products replaced with $\cstar$-products. The resulting action is
\begin{multline}
  S = \tr \Bigg\{ \int \id^4 x \bigg[ \rd^4\theta \; \e^{-gV}\dagPhi_I \e^{gV} \Phi^I 
    + \frac{1}{4} \bigg( \int \id^2\theta W^{\alpha} W_{\alpha} + \hc \bigg)
    \\
    + \I g \frac{\sqrt{2}}{3!} \bigg( \int\id^2\theta \; \epsilon_{IJK} \Phi^I [\Phi^J, \Phi^K]_{\cstar} + \int \id^2 \thetab \; \epsilon^{IJK} \dagPhi_{I} [\dagPhi_J, \dagPhi_K ]_{\cstar} \bigg)
  \bigg]
\Bigg\}
.
  \label{eq:superspaceaction}
\end{multline}
We have the standard superfields in the Wess-Zumino gauge
\begin{align}
  V& = - \theta \sigma^\mu \thetab v_\mu + \I \theta \theta \thetab \lambdab - \I \thetab\thetab\theta\lambda + \frac{1}{2} \theta\theta\thetab\thetab D
  \,,
  \\
  W_{\alpha} &= -\frac{1}{4} \bar{D}\bar{D}D_{\alpha} V
  ,
\end{align}
with the derivation operators
\begin{align}
  D_\alpha &= \frac{\partial}{\partial \theta^\alpha} +  \I \sigma_{\alpha\alphad}^\mu \thetab^{\alphad} \frac{\partial}{\partial x^\mu}
  \, ,
  \\
  \bar{D}_{\alphad} &= - \frac{\partial}{\partial \thetab^{\alphad}} - \I \theta^\alpha \sigma_{\alpha \alphad}^\mu \frac{\partial}{\partial x^\mu}
  \, ,
\end{align}
while the scalar potential is read off as
\begin{equation}
  \label{eq:scalarPotential}
  W(\Phi) = \I g \frac{\sqrt{2}}{3!}  \epsilon_{IJK} \tr \Phi^I [\Phi^J, \Phi^K]_{\cstar}
  .
\end{equation}
This obtained field theory is --- as mentioned --- still integrable and preserves $\mathcal{N}=1$ supersymmetry and conformal invariance. The original $\SU(4)_R \cong \SO(6)_R$ is broken down to a remaining $\U(1) \times \U(1) \times \U(1)$ corresponding to the original Cartan elements.  The dual theory is a type IIB string theory on $AdS_5$ times a TsT-transformed $S^5$ which mirrors the broken $R$-symmetry \cite{ci:Lunin:2005jy}.

\paragraph{The spin-chain picture of the deformed theory.}
The deformed theory still admits an integrable spin-chain interpretation of the single trace operators. However, the starred commutators in the deformed action \eqref{eq:superspaceaction} introduce $\beta$-dependent phases in some of the scalar interaction vertices. Consequently, they appear in the renormalization procedure and alter the action of the dilatation operator.

Naturally, this change enters into the spin-chain Hamiltonian but the choice of operator basis now offers two natural alternatives, as the traces can be chosen to include the $\cstar$-products or not. We stick to the basis where the operators are defined without the $\cstar$-product such that still 
\begin{align}
  \op_{\Psi} &=  \Psi^{i_1, \ldots, i_{L}} \tr \left( X_{i_1} \cdots X_{i_L} \right)
&  \longleftrightarrow &&
\Psi^{i_1 \ldots i_L}  \ket{s_{i_1}, \cdots, s_{i_L}} 
.
\end{align}
In this basis, the two-scalar-subsector dilatation operator maps to the spin-chain Hamiltonian \cite{ci:Berenstein:2004ys}
\begin{equation}
  \label{eq:betaHamiltonian}
  H^\beta = \sum_{\ell = 1}^L H^\beta_{\ell, \ell+1} 
  = \sum_{\ell=1}^L \big( 1 - 4 \; S^3_\ell S^3_{\ell+1} - 2\;  \e^{\I\beta} S^+_\ell S^-_{\ell+1} - 2\; \e^{-\I\beta} S^-_\ell S^+_{\ell+1} \big)
  ,
\end{equation}
which is diagonalized by the tweaked Bethe ansatz
\begin{equation}
  \ket{\Psi} =  \sum_{\sigma\in S^M} \sum_{1 \leq n_1 < \ldots < n_M \leq L} 
  S_{\sigma} \; \exp \left[ \I \sum_i ( p_{\sigma_i} + \beta) n_i \right] \; \ket{\vn}
  .
\end{equation}
The operator mapping condition of translational invariance now reads
\begin{equation}
  \sum_{i=1}^M (p_i + \beta) = 0
  .
\end{equation}

In the operator basis with $\cstar$-products inside the traces, the deformation parameter $\beta$ appears in the spin-chain only in the boundary conditions which is why $\beta$ is also called \emph{twist}, as mentioned in the introduction.

\section{One-point functions in the deformed theory}\label{sec:onepointFuncsInDeformed}
Having completed this short review, we seek the symmetry breaking vacuum solution in the $\beta$-deformed theory corresponding to the defect solution \eqref{eq:undefClassSol} above. We hence derive and solve the BPS equations for the scalars by setting all component fields to zero, except the complex scalars $\phi^i$, $i=1, 2, 3$, and the auxiliary fields $F^i$ and $D$. Solving for $D$ and the standard $F^i$-equations give
\begin{align}
  D &=  - \frac{g}{2} \comm{\cscal^i}{\cscalb_i}
  \,,
  &
  &\begin{cases}
    F^i = - \frac{\partial \Wb}{\partial \cscalb_i} \bigg|_{\Phi = \cscal}
    =  \I g \frac{1}{\sqrt{2}} \; \epsilon^{ijk} \starcomm{\cscalb_j}{\cscalb_k}
    ,
    \\
    \Fb_i = - \frac{\partial W}{\partial \cscal^i} \bigg|_{\Phi = \cscal}
    =   \I g \frac{1}{\sqrt{2}} \; \epsilon_{ijk} \starcomm{\cscal^j}{\cscal^k}
  \end{cases}
  .
\end{align}
Now adding boundary terms (b.t.) to the scalar Lagrangian allows the factorization
\begin{align}
  \lag_{\text{scal.}} & = \tr \Bigg\{
    -
    \bigg( \frac{1}{\sqrt{2}} \partial_\mu \cscalb_i \pm v_\mu \frac{\partial W}{\partial \cscal^i} \bigg)
    \bigg( \frac{1}{\sqrt{2}} \partial^\mu \cscal^i \pm v^\mu \frac{\partial \Wb}{\partial \cscalb_i} \bigg)
    - \frac{g^2}{8} \comm{\cscalb_i}{\cscal^i}^2
    + (\bt)
  \Bigg\}
  \label{eq:bpsFactorLag}
\end{align}
into two negative semi-definite terms whose simultaneous vanishing yields a minimum. $v^\mu$ is some real unit vector and choosing it in the $z$-direction results in the (BPS) equations
\begin{subequations}
    \label{eq:bpsEquations}
  \begin{align}
    \comm{\cscalb_i}{\cscal^i} &= 0	\,, 		\label{eq:dFlatness}
      \\
      - \frac{\partial}{\partial z} \cscal^i &= \I g \; \epsilon^{ijk} \starcomm{\cscalb_j}{\cscalb_k}
      \;.
      \label{eq:bpsFTerms}
  \end{align}
\end{subequations}
Strictly speaking, the $\cstar$-commutator is defined for the complex scalars, as the real scalars $\rscal_i$ do not carry well-defined $SU(4)_R$-charges. Nevertheless, by first expanding the $\cstar$-products into definite expressions of $q$'s, we can solve the D-flatness condition \eqref{eq:dFlatness} by simply setting $\rscal_i = 0$ for $i=4, 5, 6$. Absorbing the coupling by a field rescaling and setting
\begin{equation}
  \rscal^{\cl}_{i} = \frac{1}{z} \; t_{i} \oplus \fatZero_{(N-k)}, \qquad i = 1, 2, 3
    ,
\end{equation}
we thus get that the redefined matrices $t_i$ must satisfy the algebra
\begin{equation}
  t_i = \I  \, \epsilon^{ijk} \, [t_j, t_k]_{q}
  \label{eq:scalarAlgebra}
  .
\end{equation}
Here $[t_i, t_j]_q = q^{ij}\, t_i t_j - \frac{1}{q^{ij}} \, t_j t_i$ and $q^{ij} = q = \e^{\frac{\I}{2} \beta}$ when $(ij)$ is taken in cyclic order from $(123)$. This is the algebra of $\SU(2)_q$ in the Cartesian formulation.

A $(k\times k)$-dimensional representation satisfying the algebra \eqref{eq:scalarAlgebra} was constructed in \cite{ci:Fairlie:1989qm}, adapted here to our conventions. It is built through the standard single-entry matrices $E^i_j$ as
\begin{align*}
  t_1 &= \sum_{i=1}^{k-1} c_{k,i} \, \left( E^{i+1}_i + E^{i}_{i+1} \right),
  &
  t_2 &= \sum_{i=1}^{k-1} \I \, c_{k,i} \, \left( q^{(k-2i)} E^{i+1}_{i} - q^{-(k-2i)} E^{i}_{i+1} \right), 
  &
  t_3 = \sum_{i=1}^{k} d_{k,i} \, E^{i}_i
  ,
\end{align*}
where
\begin{align*}
  c_{k, i} &= 
  \frac{q^2 }{1 - q^4} \; \sqrt{\frac{q^2 \left(q^{4 i}-1\right) \left(q^{4 k}-q^{4 i}\right)}{\left(q^{4 i+2}+q^{2 k}\right) \left(q^{4 i}+q^{2 k+2}\right)}}
  ,
  &
  d_{k,i} &= \frac{q^2}{q^4-1} \left( q^{(k-2i+1)} - q^{-(k-2i+1)} \right)
\end{align*}
\newcommand{\hk}{\kappa}
for $i = 1, 2, \cdots, \floor{k/2} \eqdef \hk$ and
\begin{align*}
  c_{2\hk, \hk+i} &= c_{2\hk, \hk-i} , 
  & 
  &d_{2\hk, \hk+i} = - d_{2\hk,\hk-i+1}, 
  &
  &
  &
  i &= 1, 2, \cdots, \hk, 
  \\
  c_{2\hk +1, \hk+i} &= c_{2\hk+1, \hk-i+1}, 
  &
  & d_{2\hk+1, \hk+1} = 0, 
  &
  d_{2\hk+1, \hk+i+1} &= -d_{2\hk+1, \hk-i +1},
  &
  i &= 1, 2, \cdots, \hk
  .
\end{align*}

For the special case $k=2$ we simply have
\begin{equation}
  t_i = \tfrac{1}{2 \cos(\beta/2)} \; \sigma_i
\end{equation}
such that transposition gives
\begin{equation}
  \{ t_1, t_2, t_3 \}^T = \{t_1, -t_2, t_3 \}		\qquad (k=2)
  .
  \label{eq:transpositionOfTsK2}
\end{equation}
This is spoiled for $k>2$. We do, however, still have the relations
\begin{equation}
  \{ t_1, t_2, t_3 \}^T = \{t_1, - \tilde{V} t_2 \tilde{V}^{-1}, t_3 \}	
  \label{eq:transpositionOfTs}
\end{equation}
where the matrix $\tilde{V}$ is diagonal with entries $ q^{\big( 2(k-2) +6 \big) i - 2 i^2 }$, $i=1, \cdots, k$.

\subsection{Spin-chain picture of the one-point functions}
The scalar vacuum solution now allows for one-point functions which can be expressed as spin-chain overlaps just as before. The definition of the MPS \eqref{eq:defMPS} in the $\SU(2)$-subsector stays the same
\begin{equation}
  \bMPS = \tr\left( t_1 \bra{\ua} + t_2 \bra{\da} \right)^{\otimes L}
\end{equation}
but some of its properties in the spin-chain change through the deformation. 

\paragraph{Properties of the deformed MPS.}
Firstly, the requirement of both $L$ and $M$ being even for a non-zero MPS still holds true. It is easily seen from the two similarity transformations
\begin{align}
  U = U^{-1} &= \sum_{i=1}^{k} E^{i}_{k-i+1} & &\mathclap{\text{and}} & V &= V^{-1} = \sum_{i=1}^{k} (-1)^i E^i_i
  ,
\end{align}
with the property that
\begin{equation}
  U t_1 U^{-1} = t_1 , \qquad U t_{2,3} U^{-1} = - t_{2,3},
  \qquad\qquad
  V t_{1,2} V^{-1} = - t_{1,2}  , \qquad V t_{3} V^{-1} =  t_{3}
  .
\end{equation}
Inserting these into the traces in the MPS shows that the numbers of both $t_1$:s and $t_2$:s need to be even for a non-vanishing trace.

Secondly, the deformed MPS is also still parity invariant. It is immediate for $k=2$ thanks to the transposition property \eqref{eq:transpositionOfTsK2} together with $M$ being even. Numerous explicit checks suggests it to hold also for all $k$.

Importantly however, the MPS is no longer annihilated by $Q_3$ nor the other higher odd charges. In fact, the twist alters the parity property of the charges into
\begin{equation}
\label{eq:parityPropertyOfCharges}
  \parity Q_n \parity = (-1)^n \, Q_n \Big|_{\beta\to -\beta}
  .
\end{equation}
The simplifying pairing of rapidities required for a non-zero overlap before the deformation is therefore absent in the twisted spin-chain. As of now, no other replacing conditions on the rapidities have been found and the MPS has in general a finite overlap with all translationally invariant Bethe states, as is also seen in the numerical example in section \ref{sec:numerics}. 

Moreover, the twist breaks the $\SU(2)$-symmetry down to only $z$-rotations and lift the degeneracies of the $XXX_{1/2}$ spin-chain. The Bethe states are hence no longer highest weight states. This lost property was used in \cite{ci:deLeeuw:2015hxa} and also \cite{ci:Foda:2015nfk} to relate the MPS to partial N\'eel states. The determinant formula could be derived from first principles in \cite{ci:Foda:2015nfk} by writing $\kMPS = \ket{\text{N\'eel}}_M + S^- \ket{\dots}$, where the second term obviously vanishes in an overlap with a highest weight state, and constructing the N\'eel states from non-diagonal reflection matrices. This avenue appears to be blocked for a finite twist unless some other way can be found to express (the non-orthogonal part of) the MPS in building blocks satisfying the boundary Yang-Baxter equation.

Lastly, although the definition \eqref{eq:defIntegrableMPS} of integrable boundary states is out of scope here due to the relations \eqref{eq:parityPropertyOfCharges}, it is worth noting that 
\begin{equation}
  \label{eq:parityTransfParity}
  \parity \, t(u) \, \parity \, \kMPS = t(u) \kMPS  
\end{equation}
still seems to hold. It is easily proven for $k=2$ by looking at the sitewise action of the transfer matrix on the MPS\footnote{Let us ignore the complex conjugate on the $t$-matrices for notational clarity; the principle of the argument is naturally the same.}:
\begin{equation}
  \transf(u) \kMPS = \transf(u) \sum_{\vs} \tr( t_{s_1} \cdots t_{s_L} ) \ket{\vs}
  = \sum_{\vs} \tr(  \tau_{s_1} \cdots \tau_{s_L} ) \ket{\vs}
\end{equation}
where in block matrix notation
\begin{equation}
  \tau_1 = 
  \begin{pmatrix}
    t_1				&	0	\\
    \frac{\I}{u+\I} \, t_2		&	\frac{u}{u+\I} \e^{\I\beta} \, t_1
  \end{pmatrix}
  \;,
  \qquad\qquad
  \tau_2 = 
  \begin{pmatrix}
    \frac{u}{u+\I} \e^{-\I\beta} \, t_2 		& 	\frac{\I}{u+\I} \, t_1		\\
	0					&	t_2
  \end{pmatrix}
  .
\end{equation}
Making use of the transposition properties of the $t_i$:s at $k=2$, the left hand side of \eqref{eq:parityTransfParity} is
\begin{align}
  \parity \transf(u) \parity \kMPS &= \parity \transf(u) \sum_{\vs} (-1)^{\#2} \tr( t_{s_1} \cdots t_{s_L}) \ket{\vs} \notag
  \\
  &= \parity \sum_{\vs} (-1)^{\#2}\tr ( \tau_{s_1} \ldots \tau_{s_L} ) \ket{\vs}
  \notag
  \\
  &= \sum_{\vs} (-1)^{\#2} \tr ( \tau_{s_1}^T \ldots \tau_{s_L}^T ) \ket{\vs}
  .
\end{align}
The similarity transformation 
\begin{equation}
  \tilde{U} = \begin{pmatrix}
    \eye_{2\times2}		&	v t_3			\\
    v t_3		&	w \eye_{2\times 2}
  \end{pmatrix}
  ,
  \qquad v^2 = \frac{\e^{\I\beta/2} + \e^{-\I\beta/2} }{ \I + u(1+\e^{\I\beta} )}
  ,
    \qquad w^2 =  \frac{ \I + u(1+\e^{- \I\beta}) }{ \I + u(1+\e^{\I\beta}) }
    ,
\end{equation}
then mimics the behavior under transposition of the $t$:s such that 
\begin{equation}
  \tilde{U} \{ \tau_1, \tau_2 \}^T \tilde{U}^{-1} = \{ \tau_1, -\tau_2 \}
  ,
\end{equation}
from which we see that equation \eqref{eq:parityTransfParity} holds. Finding a similarity transformation for $k>2$ is not as immediate but ample numbers of explicit calculations confirm this also for larger $k$:s.

\subsubsection{Simple examples}\label{sec:simpleExamples}
While a general formula is missing, we can easily obtain the two simplest examples of the overlap, i.e. for the vacuum and for states with two excitations and comment on their large $k$-behaviours.

The same limit for the non-deformed overlap \eqref{eq:undefOverlap} was analyzed in \cite{ci:BuhlMortensen:2015gfd} and it was noted that they grow proportionally to $k^{L+1-M}$ to leading order in $k$. In particular, the vacuum overlap in the thermodynamic limit could be rearranged into an expansion in the fixed parameter $k/\sqrt{\lambda}$ which allowed for a successful comparison with the BMN-string in the D3-D5-brane system.

We find a different $k$-dependence here and a similar comparison with the dual string theory would be interesting to carry out.

\paragraph{Overlap of the vacuum.}
The operator $\tr(Z^L)$ remains protected through the deformation and its one-point function is proportional to the overlap of the MPS and the vacuum state in the spin-chain:
\begin{align}
  \braket{\MPS}{\vac} &= \tr(t_{3}^L) = \sum_{i=1}^k d_{k, i}^L
  =
  \frac{2}{(\sin \beta)^L} \sum_{i = 1}^{\floor{k/2}} \left[ \sin \tfrac{\beta ( k +1 - 2i}{2} \right]^L
    \notag
    \\
    &=
    \frac{1}{(2 \sin \beta)^L} (-1)^{\tfrac{L}{2}} \sum_{m=0}^{L} (-1)^m \binom{L}{m} \;
    \frac{\sin \tfrac{\beta k (L-2m)}{2}}{\sin \tfrac{\beta  (L-2m)}{2}}
    .
\end{align}
For computational reasons, it is easier to work with matrices $t_3$ and $t_1$ instead of $t_1$ and $t_2$, respectively and we have made this choice here thanks to the remaining $\mathbf{Z}_3$-symmetry of the scalars. 
The last equality makes it easy to read off the leading term for large $k$; the ratio of sines is of order 1  except when $\beta(L-2m) $ is a multiple of $2\pi$ at which it is of order $k$, while the binomial factor picks out the dominating term at $m = L/2$. Hence the leading part of the overlap goes as
\begin{equation}
  \braket{\MPS}{\vac} \overset{\text{large $k$}}{\sim}
  \frac{1}{(2 \sin \beta)^L} \; \binom{L}{L/2} \; k + \cdots
\end{equation}
which is notably linear in $k$.
In the limit where $\beta$ goes to zero the overlap naturally reduces to the former known case with the degree $k^{L+1}$.

\begin{figure}
  \centering
  \includegraphics[scale=1]{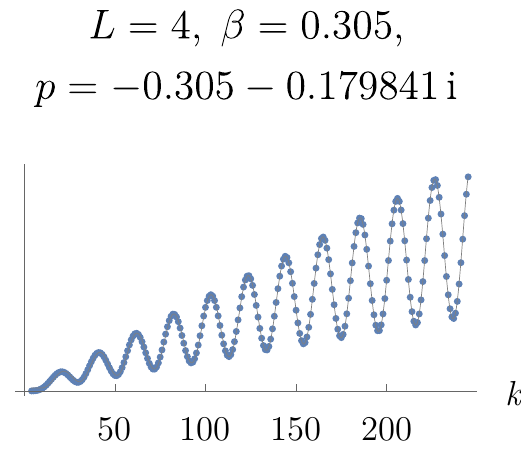}
  \includegraphics[scale=1]{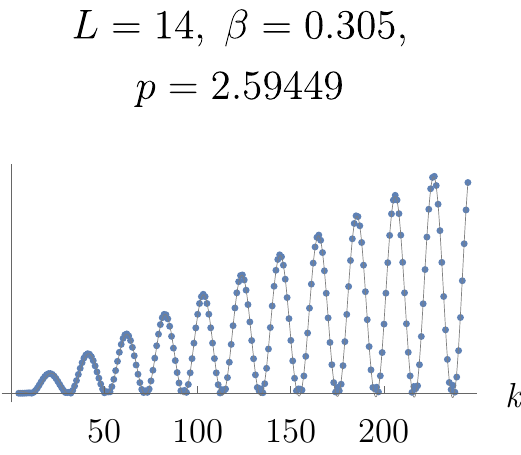}
  \includegraphics[scale=1]{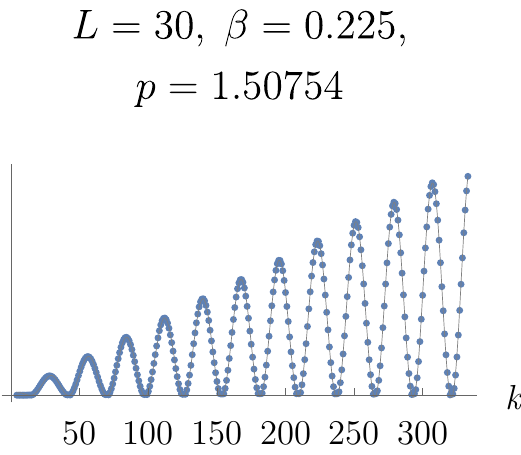}
  \caption{\label{fig:m2KDependence}The $k$-dependence of $| \braket{\MPS}{\Psi_{M=2}} |$ for three example values of $L$, $\beta$ and $p$, together with fitted ans\"atze \eqref{eq:m2Fit}.}
\end{figure}

\paragraph{The overlap of two excitations.}
 The overlap for $M=2$ is readily obtained from
\begin{equation}
  \tr(t_3^{L-2-d} t_1 t_3^d t_1) = 2 \sum_{i=1}^{k-1} c_{k, i}^2 d_{k, i}^{L-2-d} d_{k, i+1}^{d}
\end{equation}
and the translation invariance condition
\begin{equation}
  p = p_1 = -p_2 - 2 \beta
  .
\end{equation}
It reads
\newcommand{\ep}{\ensuremath{\e^{\I \tilde{p}}}}
\newcommand{\epi}{\ensuremath{\e^{-\I \tilde{p}}}}
\newcommand{\smp}{\ensuremath{\mathrm{s}_{\mp}}}
\newcommand{\epsmp}{\ensuremath{\mathrm{s}_+}}
\newcommand{\epismp}{\ensuremath{\mathrm{s}_-}}
\newcommand{\shZ}{\ensuremath{Y}}
\begin{multline}
  \braket{\MPS}{\Psi_{M=2}} =
  \frac{1}{2}  \frac{1}{(\sin\beta)^L}
  \sum_{i=1}^{k-1}  \Bigg[
    \left( 1 - \frac{\cos \beta + \cos \beta k}{\cos \beta + \cos[\beta(k-2i)]}  \right)
  \left(  \sin \tfrac{\beta(k+1 - 2i)}{2} \right)^{L-2} 
  \\
  \times \bigg(    S(p) \; \shZ_i(L, k, \beta, p) + \overline{\shZ_i(L, k, \beta, p)} \bigg)
\Bigg]
\end{multline}
where the bar denotes complex conjugation, the $S$-matrix reduces on-shell 
\newcommand{\eps}{\ensuremath{\mathrm{s}_i}}
\begin{align}
  S(p) &= - \frac{1 + \e^{2\I \beta} -2 \, \e^{-\I p}}{1 + \e^{2\I\beta} ( 1 - 2 \, \e^{\I p} ) }
  \stackrel{\text{\tiny(BAE)}}{=} \e^{-\I L(p+\beta)}
\end{align}
and the shorthand symbol $\shZ_i$ is
\begin{align}
  \shZ_i &= \e^{\I (p+\beta)} \, \frac{(L-1) - L \, \eps + (\eps)^L }{(1-\eps)^2} 
  \;,
  &
    \eps &= \e^{\I (p+\beta)} \; \frac{\sin\tfrac{\beta ( k-1 - 2 i)}{2}}{\sin\tfrac{\beta ( k+1 - 2 i)}{2}}
    .
\end{align}
The $k$-dependence is obviously more complicated than for the vacuum and there is no immediate limit behavior. However, the overlaps lie along the line in the complex plane with argument $-\tfh L(p+\beta)$ and the $k$-dependence of the absolute value is captured well by the ansatz
\begin{equation}
\label{eq:m2Fit}
  | \braket{\MPS}{\Psi_{M=2}} | \sim 
  \frac{1}{2}  \frac{1}{(\sin\beta)^L}
  \; k \big(a + b \sin(\beta k + \theta) \big)
    .
\end{equation}
The coefficients $a$, $b$ and $\theta$ have complicated dependence on $L$, $\beta$ and $p$. The ratio $a/b$ is typically in between 1 and 2, decaying overall with $L$, while $a$ has rough similarities with $L \sqrt{1 - \cos(L(p+\beta))}$. A few numeric illustrations of the $k$-dependence are plotted in figure \ref{fig:m2KDependence} together with fitted ans\"atze. Although dressed with oscillations, we note that there is still an underlying linear dependence on $k$ for $M=2$. 

\subsection[Example numerics for L=8 and M=4]{Example numerics for $L=8$ and $M=4$}\label{sec:numerics}
As mentioned in the previous section, no simplified formula for the overlaps between the MPS and Bethe states has been found nor any extra conditions on the Bethe roots as was the case without a twist. The only selection rule that remains is translational invariance and a non-zero parity invariant projection. This is confirmed by numerics and the example of $L=8$ and $M=4$ is provided here.

\begin{figure}
  \centering
  \includegraphics{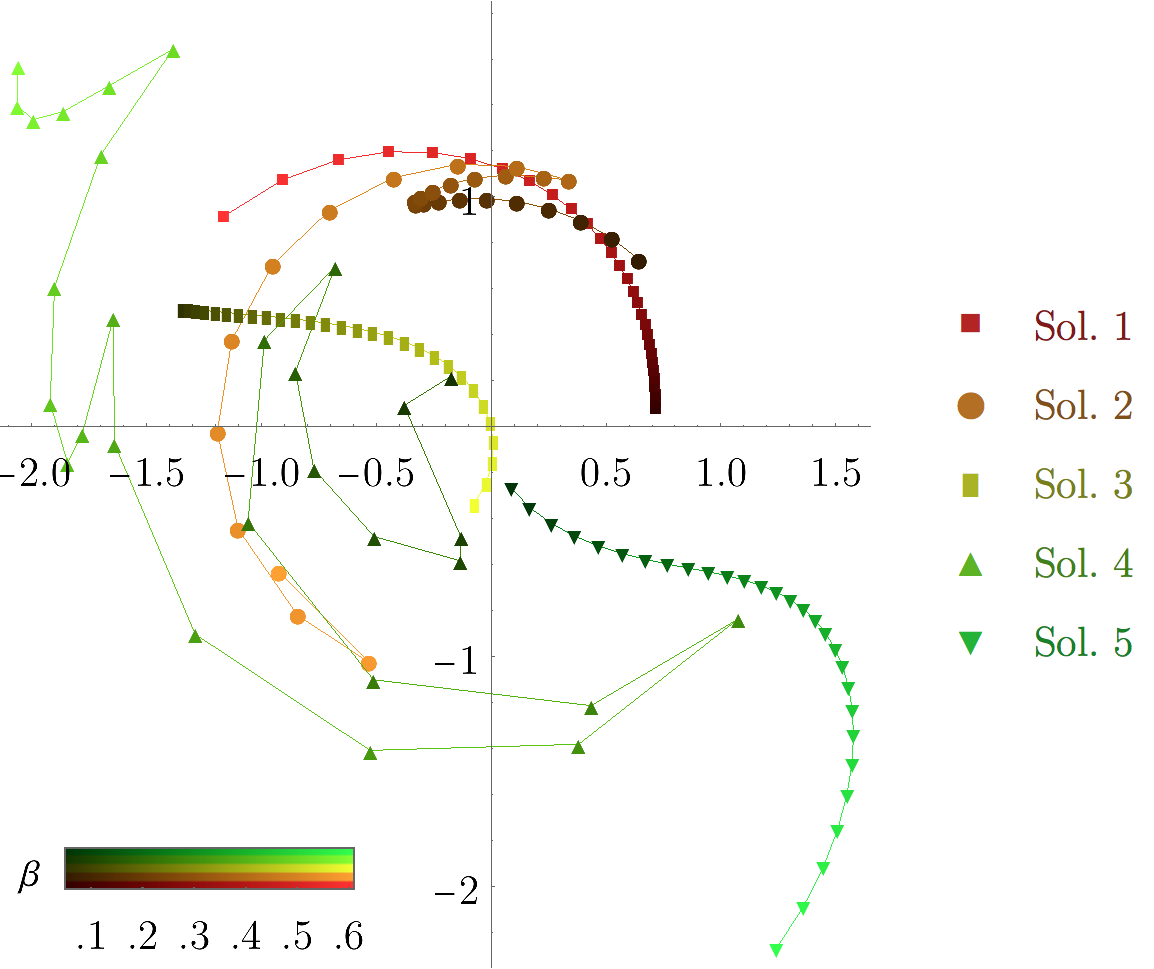}

  \vspace{3em}
  \rule{5pt}{0pt}\includegraphics{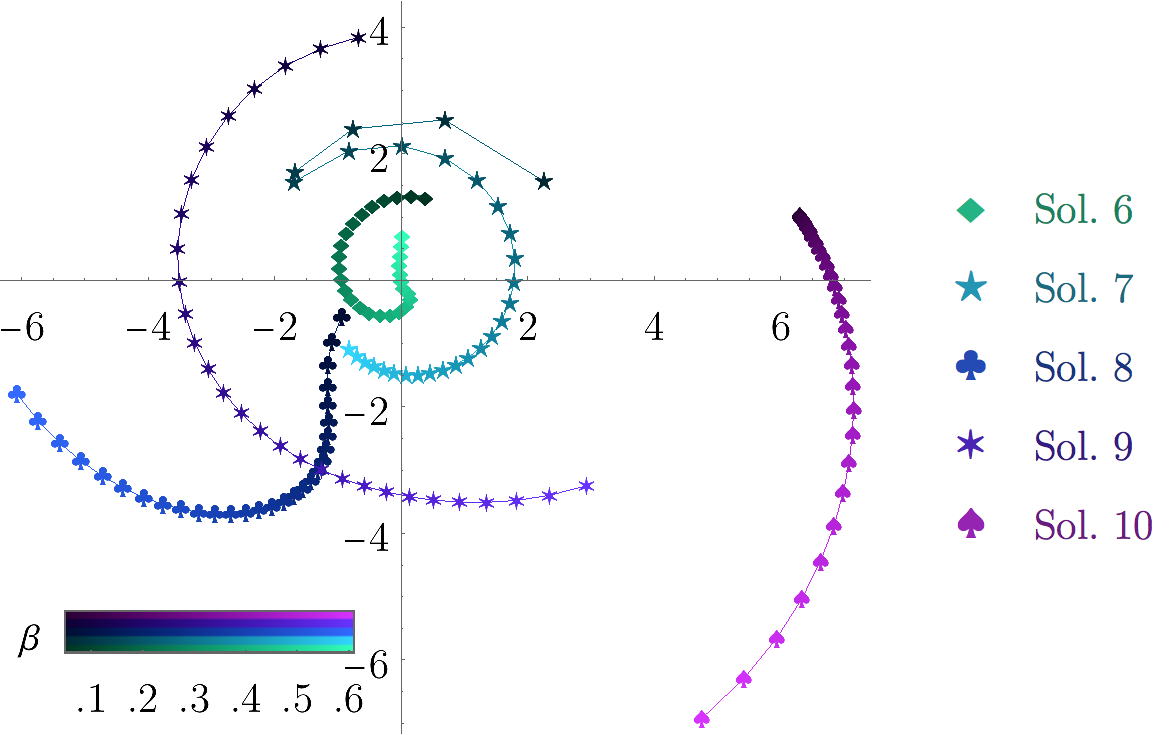}
  \caption{\label{fig:overlaps}The normalized overlaps \eqref{eq:plottedOverlaps} of the ten translationally invariant Bethe states with $L=8$ and $M=4$ plotted in the complex plane at $k=2$. Once again, the color gradients from dark to brighter colors indicate the value of $\beta$ ranging from 0.05 to 0.61.}
\end{figure}

\begin{figure}
  \centering
    \includegraphics{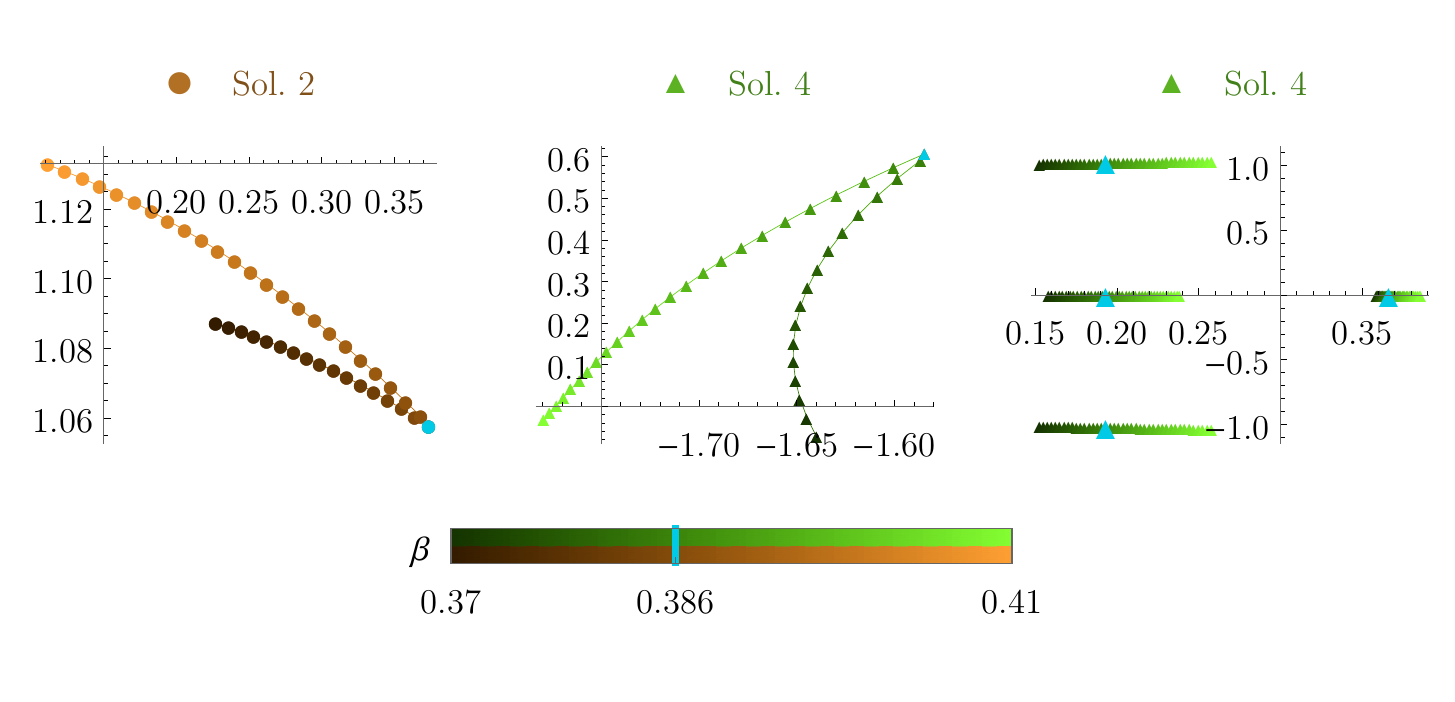}
  \caption{\label{fig:zoomedOverlaps}A magnified plot of the overlaps of solution 2 and 4 at a joint kink around $\beta \approx .386$. The kinks are due to the sudden jump in the $S$-matrices corresponding to a rapidity difference close to $-\I$, as is the case for the highlighted Bethe root solution to the right.}
\end{figure}

There are ten translationally invariant Bethe states with $L=8$ and $M=4$ and, for identification, their Bethe roots $u_i$ are plotted in the complex plane in figure \ref{fig:betheRoots} for several values of $\beta$ ranging from $0.05$ to $0.61$. The lifting of the degeneracy by $\beta$ is visible as some roots come in from infinity for the solutions corresponding to descendants at $\beta=0$. The solutions' corresponding normalized overlaps
\begin{equation}
  \label{eq:plottedOverlaps}
  2^{L-1} \frac{\braket{\MPS_{k=2}}{\Psi}}{ \sqrt{\braket{\Psi}{\Psi}} } \Bigg|_{L=8, M=4}
\end{equation}
are plotted into the complex plane in figure \ref{fig:overlaps}. They all depend rather smoothly on $\beta$ except for solutions 2, 4 and 7 which all have kinks; especially solution 4 exhibits a much more complicated pattern. The kinks occur whenever two rapidities have a difference close to $-\I$, such that the corresponding $S$-matrix has a large jump. It is clear in figure \ref{fig:betheRoots} that solution 4 has most such configuration, as is also illustrated in figure \ref{fig:zoomedOverlaps}. It shows the overlaps of solution 2 and 4 zoomed in around a simultaneous kink at $\beta\approx .386$ in juxtaposition with the Bethe roots of solution 4 at the same $\beta$-values.

\begin{figure}
  \centering
  \includegraphics{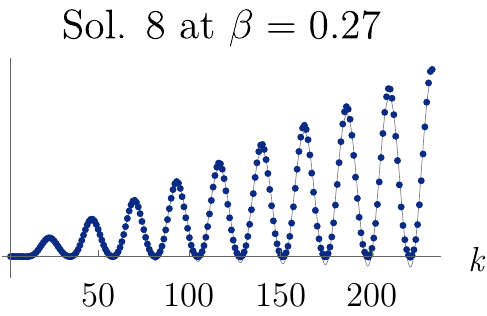}
  \includegraphics{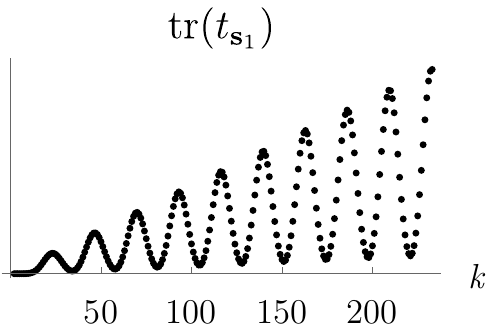}
  \includegraphics{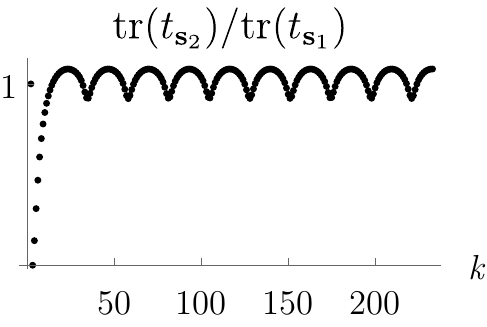}
  \caption{\label{fig:m4KDep} The $k$-dependence examplified by the overlap of solution 8 at $\beta = .27$, to the left. It is inherited from the $k$-dependence of $\tr(t_{\vs_1}) = \tr(t_3^4 \, t_1^4)$, seen in the middle, since different permutations inside the trace does not matter. The example ratio also involving $ \tr(t_{\vs_2}) = \tr( t_3^2 \, t_1 \, t_2 \, t_1^2 \, t_3 \, t_1) $ is shown to the right.}
\end{figure}

The $k$-dependence of the overlaps at $L=8$ and $M=4$ is very similar to the case at two excitations and retains the linear plus oscillatory behaviour of ansatz \eqref{eq:m2Fit}. This is due to the overall dependence on $k$ of the different trace factors in the MPS which has this form, while varying only slightly between different permutations of the matrices. The traces are all real while the component coefficients in the parity invariant part of the Bethe states all lie on a line in the complex plane. The overlap hence only restricts to this line while the absolute value is a mild variation of $\tr(t_3^4 \, t_1^4)$.  Both the $k$-dependence of an example overlap, a trace factor and the ratio between two different traces are shown in figure \ref{fig:m4KDep}. 
\newcommand{\figSpace}{\hspace{1em}}
\newcommand{\fboxgray}[1]{\fcolorbox{lightgray}{white}{#1}}
\begin{figure}
  \centering
  \includegraphics[scale=.95]{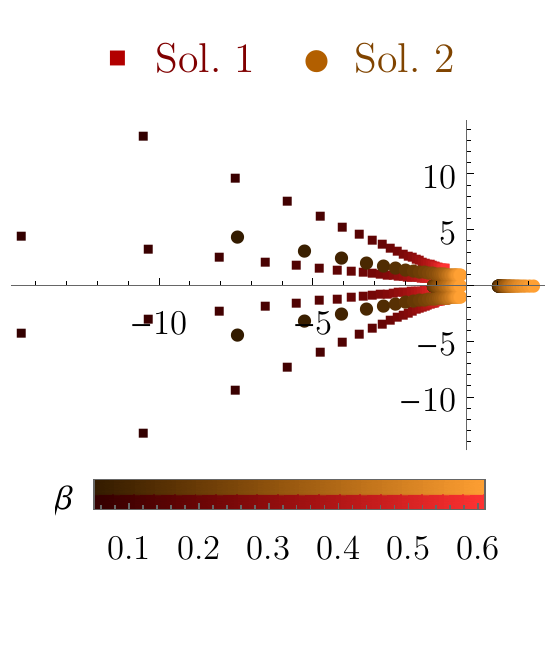}
\figSpace
    \includegraphics[scale=.95]{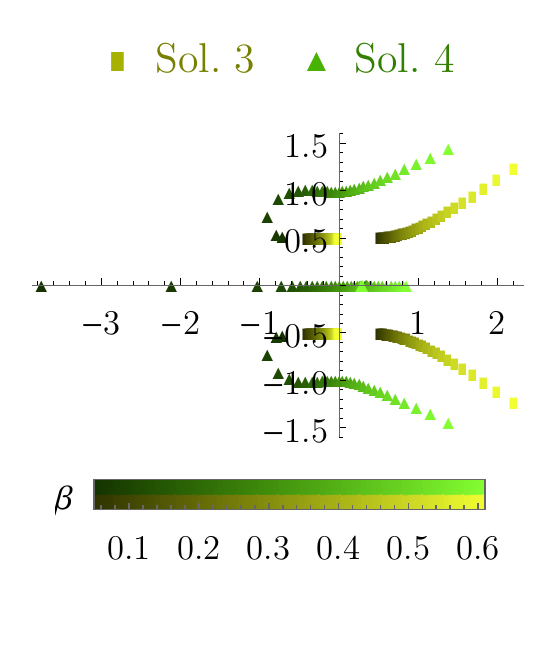}

    \includegraphics[scale=.95]{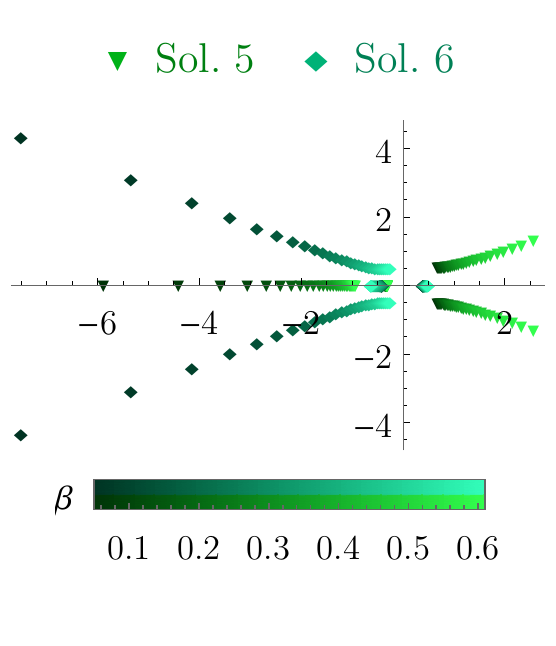}

    \includegraphics[scale=.95]{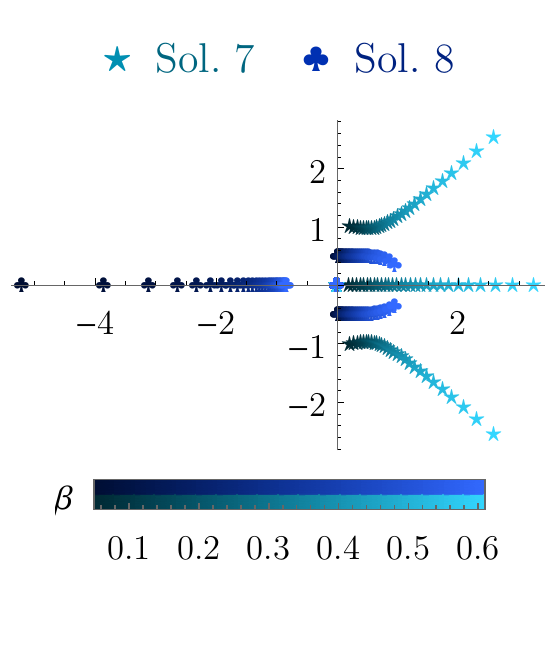}
\figSpace
    \includegraphics[scale=.95]{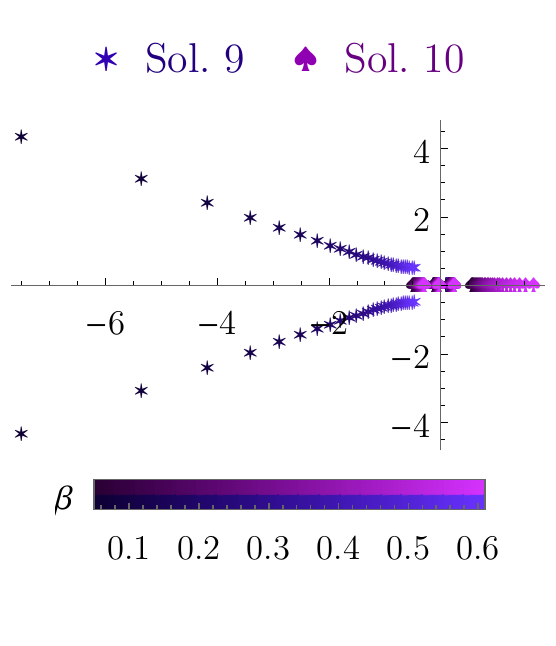}
    \caption{\label{fig:betheRoots}Plots in the complex plane of the Bethe roots $u_i$ corresponding to the ten translationally invariant states with $L=8$ and $M=4$ at various values of the twist $\beta$. The gradients from dark to brighter colors indicate which four Bethe roots that constitute the solution at the corresponding $\beta$-value in the range 0.05--0.61.}
\end{figure}
\newlength{\overlapWidth}
\setlength{\overlapWidth}{350pt}

\section{Conclusions}\label{sec:conlusions}
We have attempted to generalize earlier results for $\SU(2)$ one-point functions in $\mathcal{N}=4$ SYM with a defect to the setting of the $\beta$-deformation. While a similar classical solution for the scalars could be found, again giving the one-point functions as spin-chain overlaps with an MPS, there has been no success in reducing this expression into a compact determinant formula. The switch from $\SU(2)$ to $\SU(2)_q$ in the building blocks of the MPS retains some of its properties, such as parity invariance and the requirement of even length and number of excitations, but changes others crucial to previous constructions. Most notably, the former definition of the MPS as an integrable boundary state does no longer hold and the deformed MPS is not annihilated by odd charges. Consequently, there is no obvious pairing condition for the Bethe roots and a larger number of the states have non-zero overlap. The currently known selection rules are only 
\begin{itemize}\itemsep0pt
  \item length and numbers of excitations both even,
  \item translationally invariance and
  \item non-zero parity invariant projection.
\end{itemize}

Moreover, the breaking of the former $\SU(2)$-symmetry by a non-zero $\beta$ spoils the highest weight structures of the Hilbert space. This mars in turn previous relations made between the MPS and the N\'eel state, obstructing those earlier paths to results.

Nevertheless, analytic expressions for the overlaps were given for zero and two excitations in section \ref{sec:simpleExamples}. The former was observed to have a linear leading dependence on $k$ while the overlap with two excitations has both a linear and an oscillating part. This latter dependence carried over also to the presented numerical results for $L=8$ and $M=4$, being loosely proportional to the trace $\tr(t_1^4 \, t_2^4)$ which has such a $k$-dependence. This should be contrasted with the results at $\beta=0$ where the leading behaviour is proportional to $k^{L+1-M}$. A future comparison with the dual string theory is hence intriguing, in line with what has been done for the vacuum at $\beta = 0$ in the thermodynamic limit \cite{ci:BuhlMortensen:2015gfd}.

The ten non-zero overlaps at $k=2$, $L=8$ and $M=4$ were also presented for various values of $\beta$ together with their corresponding Bethe solutions.

As for the future prospects, there seems to be no immediate promise to find a compact determinant formula for the one-point functions in the $\beta$-deformed theory, although it might still be possible. Naturally, one could continue to study the expressions in larger subsectors but examining further the integrability properties of the MPS seems like a higher priority. The most intriguing line of inquiry might be the mentioned study of the vacuum $k$-dependence on the string theory side.

\section*{Acknowledgments}
We would like to thank R. Borsato, G. Linardopoulos, O. Ohlsson Sax and K. Zarembo for correspondence, guidance and helpful discussions and C. Marboe for his expertise and help on fast Bethe solvers.

This work was supported by the ERC advanced grant No 341222.
\makebox[0pt]{\raisebox{-4ex}[0pt][0pt]{\color{almostwhite}$\heartsuit \, 1+1=3 \, \heartsuit $}}

\bibliographystyle{elsarticle-num}

\end{document}